# Long-distance distribution of telecom time-energy entanglement generated on a silicon chip

YUAN-YUAN ZHAO,[1,2] FUYONG YUE,[1,3,4] FENG GAO,[1] QIBING WANG,[1] CHAO LI,[1] ZICHEN LIU,[1] LEI WANG,[1] ZHIXUE HE,[1] *

[1]*Peng Cheng Laboratory, Shenzhen, 518000, China*
[2]*Quantum Science Center of Guangdong-Hongkong-Macao Greater Bay Area, Shenzhen, 518045, China*
[3]*School of Microelectronics, University of Science and Technology of China, Hefei, 230026, China*
[4]*fuyongyue@ustc.edu.cn*
*\*hezhx01@pcl.ac.cn*

**Abstract:** Entanglement distribution is a critical technique that enables numerous quantum applications. Most fiber-based long-distance experiments reported to date have utilized photon pair sources generated in bulk optical crystals, with the entanglement encoded in the polarization degree of freedom. Here, we create time-energy entanglement for photon pairs generated from an on-chip silicon ring resonator via SFWM process and report the distribution of the entanglement over standard optical fiber with distance >81 km. Our work paves the way for future large-scale quantum networks with connect of distant quantum nodes.

## 1. Introduction

Quantum entanglement is a fundamental concept in quantum mechanics and is a crucial resource for various quantum communication protocols, such as entanglement based quantum key distribution [1, 2] and quantum teleportation [3]. In this application scenarios, the realization of long-distance entanglement distribution experimentally and study on the performance of entangled particles become more and more important. This process enables the establishment of secure communication channels and the interconnection of distant quantum nodes, which are essential for realizing a global quantum network [4, 5].

Compared to free-space systems, fiber-based entanglement distribution systems offer clear advantages, including low-loss propagation, environmental insensitivity, and the use of established telecommunication infrastructure and commercially available communication devices. In 2019, the distribution of polarization-entangled photon pairs over 96 km of submarine optical fiber connecting Malta and Sicily was reported, where >90% polarization visibility is achieved with photon pair rate of 257 Hz [6]. In 2022, polarization-entangled photon pairs distributing over international link between Austria and Slovakia with distance of 248 km was demonstrated, with stable detection photon pair rates of 9 Hz over 110 hours [7]. One year later, another international fiber link with 224 km submarine fiber between United Kingdom and the Republic of Ireland has been tested for entangled photon distribution [8]. In 2024, researchers have successfully demonstrated quantum network protocols operating over urban fiber optic links spanning distances of up to 14 kilometers partially overheaded and partially underground [9]. These successful experiments showcase the great progress in this critical area of quantum technology, which is essential for the advancement of quantum communication and the eventual realization of a functional quantum internet.

While many experiments have been reported in distributing entanglement over long-distance fiber links, the majority of fiber-based experiments reported to date have utilized photon pair sources generated in bulk optical crystals, with the entanglement encoded in the polarization degree of freedom [6, 10-13]. These setups often require complicated compensation schemes to stabilize the polarization drift in long fibers [14]. Energy-time-entangled photon pairs at telecom band, on the other hand, offer a simple and good-enough alternative for fiber-based long distance quantum distribution due to their insensitivity to environmental influences on the transmission link [15].

Chip-integrated entangled photon pair sources [16-18], based on spontaneous four-wave mixing (SFWM) in the telecommunications band, have been proved to be promising in spectral brightness, channel number, photon purity, and the potential for ultra-compact integration. On-chip sources of entangled pairs have been integrated with various photonic platforms, including silicon [19], silicon nitrides [20], indium phosphide [21], gallium nitride [22], aluminum gallium arsenide [23], and lithium niobate [24]. Notably, Silicon-chip-based photon pair source is a versatile platform owing to its complementary metal-oxide-semiconductor (CMOS) compatibility, relative low waveguide loss and high nonlinearity compared to glass and silicon nitride. Specifically, entangled photon pair source based on resonance-enhanced silicon micro-ring benefits from high quality factor, small mode volume size, flexible free spectral range (FSR), and multiple wavelength channels [19].

While making great progress in silicon-chip based photon pair sources in recent years, the performance of entanglement distribution of silicon-chip-based entangled photon pair source in the scenario of long-distance fiber link has been rarely explored. In this work, we create time-energy entanglement for correlated photon pairs generated from an on-chip silicon ridge racetrack ring resonator via the SFWM process and report the distribution of the entanglement over standard optical fiber with fiber length longer than 81 km. The measured Coincidences-to-Accidentals Ratio (CAR) and second order auto-correlation function of photon pairs generated from the on-chip silicon ring are CAR = 5218±470 at a pump power of 0.198 mW and $g^{(2)}(0)$ = 2.073±0.09, indicating the high performance and suitability of the on-chip silicon ring for generating photon pairs towards applications of long-distance entanglement distribution. The measured visibility of time-energy entanglement of the generated photon pairs before and after 81 km single mode fiber propagation is 89.5±0.9% and 88.6±2.5%, respectively. Our demonstration of long-distance entanglement distribution from an on-chip silicon micro-ring photon pair source with high brightness has the potential to serve as an important resource in quantum communication and quantum internet.

## 2. Results

### 2.1 Silicon chip design and experimental setup

As shown in Fig. 1(a), a third-order nonlinear process of degenerate SFWM is used to produce photon pairs on the silicon chip. In the process, two photons from a single pump annihilate and create the signal and idler photons at new frequencies equally spaced on either side of the pump frequency. Each SFWM process conserves the momentum and energy of the two input photons. SFWM occurs in the optical cavities formed by the ridge racetrack ring resonator.

The silicon chip is fabricated using standard 220 nm silicon-on-isolator (SOI) Foundry processes with 193-nm lithography. The width of the waveguide, thickness of ridge and slab waveguide are w=500 nm, h1=105 nm, h2=115 nm, respectively, as shown in the inset of Fig. 1(a). The length of racetrack is designed as 400 $\mu$m, and the radius of the ring is 50 $\mu$m. The ridge racetrack ring is evanescently coupled with the bus silicon ridge waveguide via point coupler with gap of 450 nm. The bus waveguide ends with inverse tapers with length of 500 um and tip size of 200 nm, designed for input/output lensed fiber coupling. A micro-heater is added above one racetrack to facilitate electro-optic control and tune the resonant wavelength of the microring resonator. A microscopy image of the chip is shown in the inset in Fig. 1(b).

The chip is mounted on a temperature-controlled stage with a thermal feedback system consisting of a thermo-electric controller and a thermistor on the stage mount to stabilize the resonance and photon emission. Lensed fibers with spot size of 3.3 $\mu$m are used to couple light into and out of the chip through the spot size converters (SSCs). Neglecting the propagation loss inside the straight waveguide, the measured insertion loss of each fiber-to-waveguide SSC is about 2.5 dB over the telecom wavelength of interest. In the transmission measurement, a continuously tunable external cavity laser (spectral bandwidth < 10 kHz) with input powers of 0.1 dBm is used to characterize the transmission spectra from 1460 nm to 1640 nm.

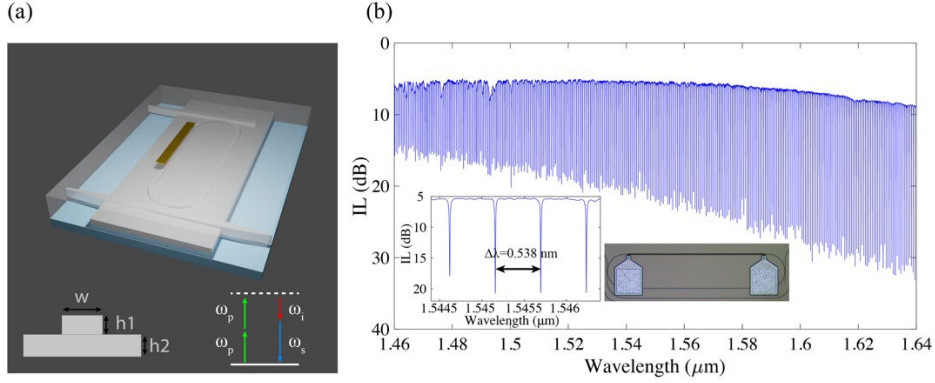

Fig. 1. (a) Schematic of the silicon chip. The ridge racetrack ring is evanescently coupled with the bus silicon ridge waveguide via point coupler with gap of 450 nm. The width of the waveguide, thickness of ridge and slab waveguide are w=500 nm, h1=105 nm, h2=115 nm, respectively, as shown in the inset. The bus waveguide ends with inverse tapers with length of 500 um and tip size of 200 nm, designed for input/output lensed fiber coupling. A metal strip (width=2 $\mu$m, length= 400 $\mu$m) is placed above the race waveguide spaced with SiO$_2$ layer for tuning the resonant wavelength through ohmic heating. (b) Measured transmittance of the silicon chip. An optical microscope image of the ring is shown in the inset. The FSR at $\lambda$=1.545 nm is 67 GHz ($\Delta\lambda$~ 0.538 nm). The quality factors of the four resonant wavelengths shown in the inset are 0.96×10$^5$, 1.01×10$^5$, 1.07×10$^5$, 1.07×10$^5$, respectively.

The designed free spectral range (FSR) of the silicon micro-ring resonator is 67 GHz. This small FSR (narrower frequency-mode spacing), can provide more wavelength pairs using wavelength demultiplexing technology, which is helpful in multiuser networks [25]. Fig. 1(b) shows the measured Q-factors of four selected resonant wavelengths ($\lambda_1$=1544.62 nm, $\lambda_2$=1545.16 nm, $\lambda_3$=1545.7 nm, $\lambda_4$=1546.23 nm) around the pump wavelength are $Q_1$=0.96×10$^5$, $Q_2$=1.01×10$^5$, $Q_3$=1.07×10$^5$, $Q_4$=1.07×10$^5$, respectively. These Q-factors make a good balance between stability and brightness, since high Q-factor may bring higher photon pair generation rate (PGR) but make it challenging to maintain on-resonance performance and cause low clock rate in the sequential Time-Bin entanglement scheme due to the long coherence time and narrow bandwidth [26].

The experimental setup for generation and characterization time-energy entangled photon pairs is shown schematically in Fig. 2. The chip is pumped by laser with wavelength of $\lambda_p$ =1545.69 nm, corresponding to International Telecommunication Union (ITU) channel 39.5. Two 50 GHz Wavelength Division Multiplexing (WDM) filters with total isolation of 120 dB are used to purify the pump light and attenuate the background noise. A variable optical attenuator (VOA) and a 95:5 beam splitter (BS) are inserted in the setup for the purpose of adjusting and monitoring the laser power entering the chip. Band-stop filters (BSFs) are used to remove the pump light from the output. A polarization controller (PC) is used to control the polarization of the pump light. A 50:50 BS is employed to separate the signal and idler photons into two distinct paths. Subsequently, a tunable filter (TF) is placed in each path to select and extract the signal ($\lambda_s$ = 1551.66 nm) and idler ($\lambda_i$ = 1539.80 nm) photons. The laser source, PC, BSFs, tunable filters, BSs are all fiber integrated, off the chip. The generated photon pairs are then detected by two superconducting nanowire single-photon detectors (SNSPDs) with a detection efficiency of about 80% (dark count rate <100 Hz, time resolution < 15 ps). The output signals from the SNSPD are analyzed by a time correlator (IDQ800). The signal photons are transmitted through an 81 km single mode fiber. An asymmetric Mach-Zehnder interferometers (AMZI), with 1.5 ns time delay between short and long paths, is utilized to split the input pulse into two identical parts that define two time-bins (early and late) for creating time-energy entanglement.

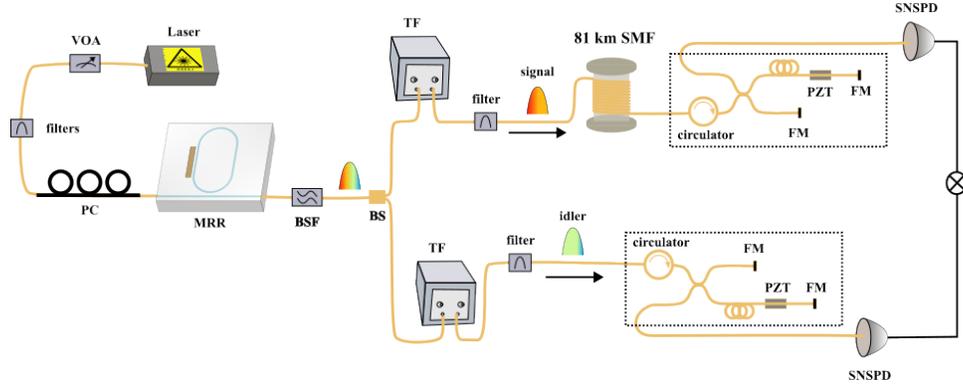

Fig. 2. Experimental setup. The filtered continuous-wave (CW) laser is injected into the silicon MRR to generate pairs of signal and idler photons through the SFWM process. The signal photon is then transmitted over single-mode fiber (SMF) with a distance of 81 km. The entanglement properties are analyzed using Franson interferometry. Abbreviations: VOA - variable optical attenuator; PC - polarization controller; BSF – bandstop filter; BS - beam splitter; TF - tunable filter; SMF - single-mode fiber; PZT - piezoelectric ceramics; FM - Faraday mirror; SNSPD - superconducting nanowire single-photon detectors. Further details are provided in the main texts.

*2.2 Characterizations of entangled photon source*

To characterize the performance of the on-chip photon pair source, several key parameters are evaluated and analyzed: single counting rate ($C_{s,i}$), coincidence counting rate ($C_c$), CAR, spectral purity, and energy-time entanglement.

**Single counting rate**

The SFWM process in micro-ring is determined by several characteristics that enable the efficient generation of correlated photon pairs. The generation rate can be calculated by [16, 17]: $R = (L\gamma)^2 F^6 \frac{v_g}{2L} P_p^2$, where L is the ring circumference, $\gamma$ is the nonlinear coefficient, $F = \sqrt{\frac{2Qv_g}{\omega_0 L}}$ is the on-resonance field enhancement in the microring [17], $v_g$ is the group velocity, Q is the quality factor, $P_p$ is the pump power. The correlated photon pairs generated through SFWM process depend quadratically on the power of the pump. This quadratic dependence helps distinguish the photon pairs of interest from photonic noises, such as those from the spontaneous Raman process, which scale linearly with the pump power.

Fig. 3(a) shows the single side counting rate at different pump powers. By fitting the experimental data with a function of the form $C_s = aP^2 + bP$, we obtained the coefficients $a = 3.4 \times 10^4$ Hz/mW$^2$, and $b = 1.9 \times 10^4$ Hz/mW, respectively. The results indicate that the signal photons dominate over the linear photonic noise.

**Coincidence counting rate**

By acquiring the coincidence histogram between the signal and idler detections, we could observe a coincidence peak with zero delay. The measured coincidence counts $C_c$ with 1 ns time bin as a function of the on-chip pump power and the corresponding quadratic fitting are shown in Fig. 3(b), indicating that the collected photon pairs are generated through the expected SFWM process in the MRR, the quadratic fitting coefficient is 545.5 Hz/mW$^2$. The PGR is calculated as $C'_s C'_i / C_c$, where $C'_{s(i)}$ is the single side counting rate from SPWM, and hence the quadratic terms obtained from the fitting curve in Fig. 3(a). Therefore, the PGR is estimated to be $2.1 \times 10^6$ pairs.s$^{-1}$.mW$^{-2}$. Taking into account the resonance linewidth of the MRR (~ 2 GHz), the resultant brightness of the photon pair source is about $1 \times 10^6$ pairs.s$^{-1}$.GHz$^{-1}$.mW$^{-2}$.

**Coincidence-to-Accidental Ratio (CAR)**

The CAR is another important figure of merit for a correlated photon source, which can be expressed as $\text{CAR} = \frac{C_c}{C_a}$, where $C_a$ are the measured accidental coincidence counts. In our experiment, the bin width is set as $\tau = 20$ ps. Neglecting the influence of the detector dark counts, $\text{CAR} \propto \eta_s \eta_i / C_c \tau$, where $\eta_s$ and $\eta_i$ are the efficiencies of the signal and idler channels, respectively [27]. CAR is calculated by integrating the coincidence counts within the FWHM (205 ps) time window and dividing by $C_a$ within the same time interval. In Fig. 3(b), the data are obtained by collecting photons in 60 seconds. At high pump power, the reduction of CAR is mainly due to multi-photon emission, while at lower pump power, the CAR is limited by the detector's dark counts. For a pump power of 0.198 mW, we measure a CAR of 5218±470.

**Spectral Purity**

The second-order auto-correlation function $g^{(2)}(0)$ is a key parameter for evaluating spectral purity, which is essential for quantum teleportation and other multi-photon schemes. In experiment, $g^{(2)}(0)$ can be measured by sending the signal or idler photons into a balanced 50:50 beam-splitter and measuring the coincidence counts between its two outputs. By fitting with a Lorentzian function, the obtained $g^{(2)}(0)$ is $2.073 \pm 0.09$ (Fig. 3(c)), which is close to the ideal case of single mode $g^{(2)}(0) = 2$. Our results indicate that the performance of our on-chip photon source makes it suitable for the applications of entanglement-based quantum networks.

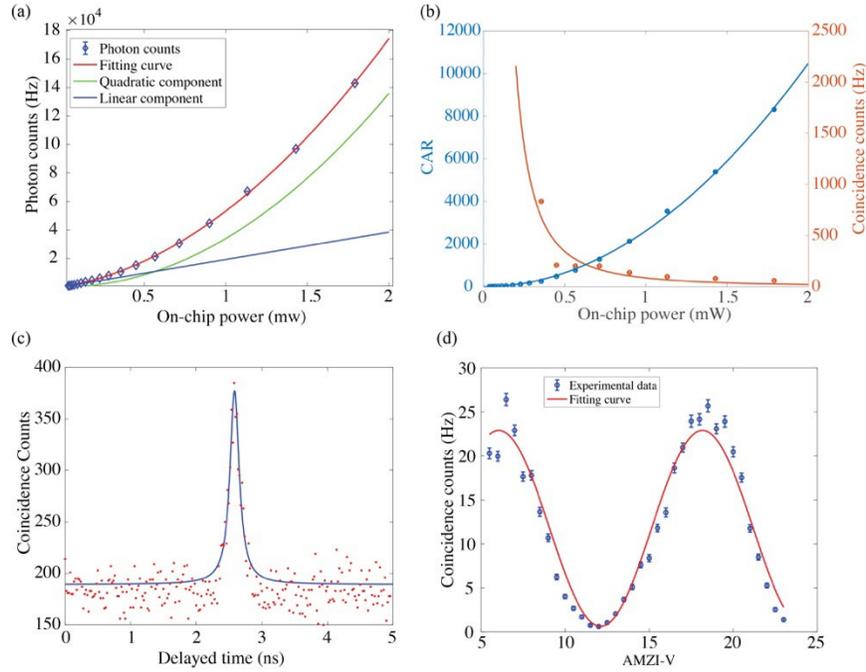

Fig. 3. (a) Measured photon counts as a function of injected laser power on the chip. (b) CAR and coincidence. (c) Measured $g^{(2)}(0)$, the counts are obtained by integrating 600 s. (d) Measured energy-time entanglement visibility.

**Energy-Time Entanglement**

Next, we proceed to the characterization of energy-time entanglement. This entanglement can be verified using AMZI. In our work, two separate fiber-based AMZIs are used for the signal photon and idler photon, respectively. Our AMZIs are constructed entirely from fiber components and are enclosed in a box to reduce external disturbances. The time delay between the two arms is designed to be 1.5 ns, which is larger than the coherence time of the single photon. A PZT is used to vary the relative phase between the two AMZIs. Fig. 3(d) shows the experimentally measured data and fitted curve with a sinusoidal function. The normalized coincidence probability is given by $\frac{1}{2}(1 + \cos(\phi_s + \phi_i))$, where $\phi_s$ and $\phi_i$ are the relative phase introduced at the long arms of the two AMZIs. In our experiment, we varied $\phi_s$ by adding voltage from 5.5 V to 23 V, keeping $\phi_i$ unchanged. We observe a visibility of 95.4±7% Note that for each phase step, the data are obtained by averaging the coincidence counts over 60 seconds with a pump power of 0.283 mW.

### 2.3 Long distance entanglement distribution

The results demonstrated above indicate the well performance of our on-chip photon pair source in various figures of merits. Next, we proceed to investigate the long-distance entanglement distribution in single mode fiber with this on-chip source.

In our experiment, the measured spectral full width at half maximum (FWHM) of the MRR is about $\delta\lambda$ =0.0176 nm, which corresponds to a frequency bandwidth of 2.2 GHz. Assuming a transform-limited Gaussian pulse, the width of the time bin peak is $t_0 \approx 205.5$ ps. The fiber employed in our experimental setup is G.652D single mode fiber, with a dispersion parameter D≈18 ps/nm.km. After transmission over a distance of z≈81 km, the pulse will experience an additional temporal broadening that can be obtained as $\Delta t \approx D\delta\lambda z$=25.7 ps. Consequently, the pulse broadening caused by chromatic dispersion is expected to be $\sqrt{t_0^2 + \Delta t^2} = 207.1$ ps. The actual pulse width observed in the experiment is about 206.6 ps, which agrees well with the theoretical estimation, as shown in Fig. 4(a) and (b).

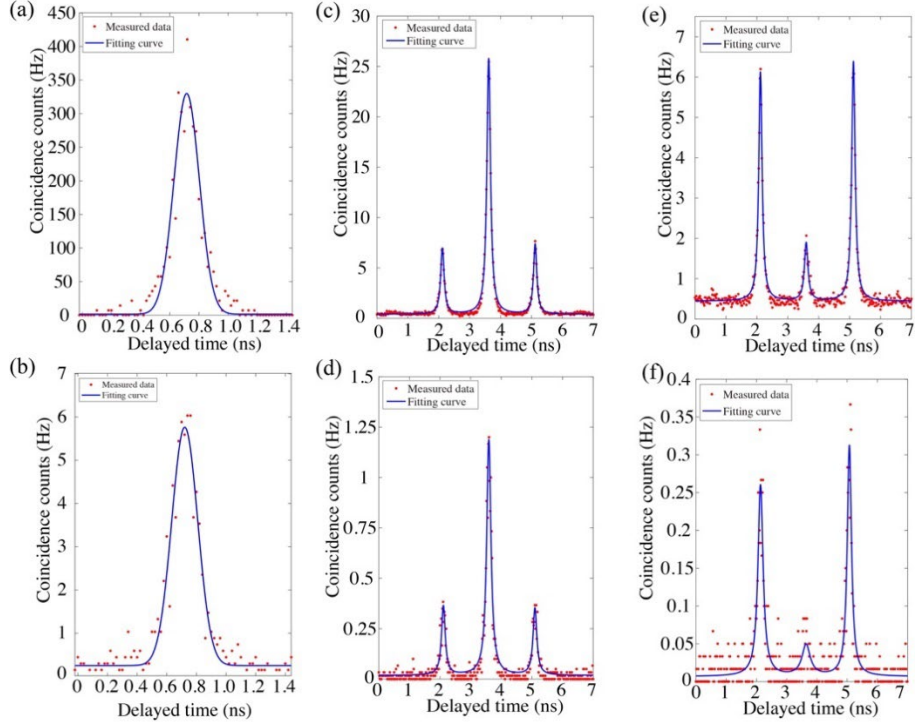

Fig. 4. Coincidence counts (a) before and (b) after the 81 km SMF transmission. Coincidence counts at maximum constructive interference (c) before and (d) after the SMF, and at destructive interference (e) before and (f) after the SMF. The discretized distribution of the coincidence counts in (f) are mainly coming from the discretization of photon counting.

The loss of the fiber is calculated by comparing photon counts with the same integration time before and after the 81 km fiber transmission. This measurement shows that the fiber transmission loss results in a 16.8 dB reduction in the photon count. For large pump powers, more photons can be detected per second. However, multi-photon effects become significant and tend to decrease the visibility of the energy-time entanglement. With an on-chip pump power of 2.27 mW, we observe an energy-time entanglement of about 89.5±0.9%. Fig. 4(c) and (e) illustrate the maximum and minimum coincidence peaks. The gap between two adjacent peaks equals the time delay between the two arms of the AMZI. After transmission through an 81 km single-mode fiber, the visibility becomes 88.6±2.5%, as shown in Fig. 4(d) and (f). Notably, energy-time entanglement proves highly resilient to transmission noise in fiber-based long-distance communication. The data are obtained by integrating over a 60-second period. Here, due to ambient temperature oscillations, the coincidence counts during the collection time slightly deviate from the maximal/minimum interference cases. Future improvements can be achieved by shielding the AMZI in a temperature-stable environment, which only involves the local setups and avoid the need for complex noise compensation systems in long-distance fiber links [14, 28, 29].

## 3. Discussion and Conclusion

Our on-chip photon pair source exhibits great performance in terms of noise reduction, spectral purity, and entanglement visibility, making it a promising candidate for scalable, fiber-based quantum communication networks. Specifically, the high CAR value observed in the experiment indicates the low noise, which is crucial for practical quantum communication systems. The measured $g^{(2)}(0)$ approximates the ideal case for a single mode, confirming the suitability of our source for multi-photon applications such as quantum teleportation. Our investigation into long-distance entanglement distribution reveals that the photon source maintains high visibility of energy-time entanglement, with only minimal pulse broadening due to the chromatic dispersion of the fiber. This resilience to transmission noise paves the way for integrating the on-chip source with existing fiber communication infrastructures in practical quantum applications.

**Funding.** This work is support by the Major Key Project of PCL, Guangdong Provincial Quantum Science Strategic Initiative (GDZX2303001 and GDZX2200001), and the Innovation Program for Quantum Science and Technology (No. 2021ZD030290001).

**Acknowledgments.** We thank Perry Ping Shum from Southern University of Science and Technology for help on silicon chip fabrication.

**Disclosures.** The authors declare no conflicts of interest.

**Data availability.** Data underlying the results presented in this paper are not publicly available at this time but may be obtained from the authors upon reasonable request.